\documentclass[10pt,letterpaper]{article}
\usepackage[top=0.85in,left=2.75in,footskip=0.75in]{geometry}

\usepackage{changepage}

\usepackage[utf8]{inputenc}

\usepackage{textcomp,marvosym}

\usepackage{fixltx2e}

\usepackage{amsmath,amssymb}

\usepackage{cite}

\usepackage{nameref,hyperref}

\usepackage{microtype}
\DisableLigatures[f]{encoding = *, family = * }

\usepackage{rotating}

\usepackage{multirow,booktabs}
\usepackage{amsmath,amssymb,latexsym,MnSymbol}
\usepackage{hyperref}

\raggedright
\usepackage[aboveskip=1pt,labelfont=bf,labelsep=period,justification=raggedright,singlelinecheck=off]{caption}

\makeatletter
\renewcommand{\@biblabel}[1]{\quad#1.}
\makeatother

\date{}

\usepackage{lastpage,fancyhdr,graphicx}
\usepackage{epstopdf}
\fancyheadoffset[L]{2.25in}
\fancyfootoffset[L]{2.25in}
\begin{document}
\vspace*{0.35in}

\begin{flushleft}
{\Large
\textbf\newline{Scale-adjusted metrics for predicting the evolution of urban indicators and quantifying the performance of cities}
}
\newline
\\
Luiz G. A. Alves\textsuperscript{1,2},
Renio S. Mendes\textsuperscript{1,2},
Ervin K. Lenzi\textsuperscript{1,2,3},
Haroldo V. Ribeiro\textsuperscript{1,4,*},
\\
\bigskip
\bf{1} Departamento de F\'isica, Universidade Estadual de Maring\'a, Maring\'a, PR 87020-900, Brazil
\\
\bf{2} National Institute of Science and Technology for Complex Systems, CNPq, Rio de Janeiro, RJ 22290-180, Brazil
\\
\bf{3} Departamento de F\'isica, Universidade Estadual de Ponta Grossa, Ponta Grossa, PR 84030-900, Brazil
\\
\bf{4} Departamento de F\'isica, Universidade Tecnol\'ogica Federal do Paran\'a, Apucarana, PR 86812-460, Brazil
\\
\bigskip

%
%





* hvr@dfi.uem.br

\end{flushleft}
\section*{Abstract}
More than a half of world population is now living in cities and this number is expected to be two-thirds by 2050. Fostered by the relevancy of a scientific characterization of cities and for the availability of an unprecedented amount of data, academics have recently immersed in this topic and one of the most striking and universal finding was the discovery of robust allometric scaling laws between several urban indicators and the population size. Despite that, most governmental reports and several academic works still ignore these nonlinearities by often analyzing the raw or the per capita value of urban indicators, a practice that actually makes the urban metrics biased towards  small or large cities depending on whether we have super or sublinear allometries. By following the ideas of Bettencourt \textit{et al.} [PLoS ONE 5 (2010) e13541], we account for this bias by evaluating the difference between the actual value of an urban indicator and the value expected by the allometry with the population size. We show that this scale-adjusted metric provides a more appropriate/informative summary of the evolution of urban indicators and reveals patterns that do not appear in the evolution of per capita values of indicators obtained from Brazilian cities. We also show that these scale-adjusted metrics are strongly correlated with their past values by a linear correspondence and that they also display crosscorrelations among themselves. Simple linear models account for 31\%-97\% of the observed variance in data and correctly reproduce the average of the scale-adjusted metric when grouping the cities in above and below the allometric laws. We further employ these models to forecast future values of urban indicators and, by visualizing the predicted changes, we verify the emergence of spatial clusters characterized by regions of the Brazilian territory where we expect an increase or a decrease in the values of urban indicators.



\section*{Introduction}
Over the past six decades the world went through a period of quick and remarkable urbanization. According to the United Nations~\cite{UN} in the year of 2007, for the first time, the world urban population has surpassed the rural one and, if this process persists, two-thirds of the world population are expected to be living in urban areas by 2050. On one hand, cities are usually associated with higher levels of literacy, health care and better opportunities; on the other hand, unplanned urbanization and bad political decisions also lead to pollution, environmental degradation, growth of crime, unequal opportunities and the increase in the number of people living in substandard conditions. In this sense, there is a vast necessity for finding patterns, quantifying and predicting the evolution of urban indicators, since these investigations may provide guidance for better political decisions and resources allocation. 

Fostered by this need and also due the availability of an unprecedented amount of data at city level, several researchers have recently promoted an impressive progress into what has been called \textit{Science of Cities}~\cite{Louf}. These new data allowed researchers to probe patterns of the cities to a degree not before possible and one of the most striking and universal finding of these studies was the discovery of robust allometric scaling laws between several urban indicators and the population size. Patents, gasoline stations, gross domestic product~\cite{Bettencourt,Arbesman,Bettencourt2,Bettencourt3}, crime~\cite{Bettencourt3,Gomez-Lievano,Alves,Alves2,Alves3,Ignazzi}, indicators of education~\cite{Ignazzi}, suicides~\cite{Melo}, number of election candidates~\cite{Mantovani,Mantovani2}, transportation networks~\cite{Samaniego,Louf2}, employees from several sectors~\cite{Pumain}, measures of social interaction~\cite{Pan} are just a few examples where robust scaling laws have been found. Despite these intrinsic nonlinear relationships it is common to find works that try unveiling relationships between population size and urban indicators by employing linear regressions in the raw data; also, per capita indicators  (that is, divided by population size) are ubiquitous in reports of government agencies and are often used as a guide for public policies when analyzing the temporal evolution of a given city or for comparing/ranking the performance of cities with different population sizes. However, these linear regressions may result in misguided/controversial relationships~\cite{Gordon,Alves2} and per capita indicators are oblivious to the allometric scaling laws that make city a complex agglomeration that cannot be modeled as a linear combination of its individual components~\cite{Bettencourt3,Bettencourt4,Pumain2}. In this sense, there is a paucity of alternative metrics for urban indicators that may overcome the foregoing problems and provide a fairer comparison between cities of different sizes as well as a better understanding of the city evolution. Bettencourt \textit{et al.}~\cite{Bettencourt3} have recently proposed a simple alternative to overcome these nonlinearities by evaluating the difference between the actual value of the urban indicators and the value expected by the allometries with the population size (that is, the residuals in the allometric relationships). This scale-adjusted metric explicitly considers the allometric relationships and already have proved to be useful in the economic context~\cite{Lobo,Podobnik} and for unveiling relationships between crime and urban metrics that are not properly carried out by regression analysis~\cite{Alves2}. 

In this article, we follow the evolution of this relative metric for eight urban indicators from Brazilian cities in three years (1991, 2000 and 2010) in which the national census took place. By grouping cities in above and below the allometric laws, we argue that the average of this scale-adjusted metric provides a more appropriate/informative summary of the evolution of the urban indicators when compared with the per capita values; it also reveals patterns that do not appear when analyzing only the evolution of the per capita values. For instance, while the per capita values of homicides have systematically increased over the last three decades, both the average of the scale-adjusted metric for cities above and below the allometric law are approaching zero, that is, cities where the number of homicides is above the expected by the allometry have managed to reduce this crime, whereas this crime has been increased in those cities where number of homicides is below the allometry. We argue that the nonlinearities may affect the per capita indicators by creating a bias towards large cities for superlinear allometries and a bias towards small cities for sublinear allometries. We further show that these scale-adjusted metrics are strongly correlated with their past values by a linear correspondence, making them particularly good for predicting future values of urban indicators through linear regressions. We have tested this hypothesis via a linear model where the scale-adjusted metric for one indicator in a given census was predicted by a linear combination of all eight metrics evaluated from the preceding census. These simple models account for 31\%-97\% of the observed variance in data and correctly reproduce the average value of the scaled-adjusted metric when grouping the cities in above and below the allometric laws. Motivated by these good agreements, we present a prediction for the values of urban indicators in the year of 2020 by assuming the linear coefficients constants over time. By visualizing the predicted changes, we verify the emergence of spatial clusters characterized by regions of the Brazilian territory where the model predicts an increase or a decrease in the values of urban indicators. We further report a list containing all the scale-adjusted metrics as well as the predictions for each city in the hope that government agencies find these informations useful (\nameref{S1_dataset}).

\section*{Results and Discussion}
\begin{figure}[!ht]
\begin{adjustwidth}{-2.25in}{0in}
\begin{center}
\includegraphics[scale=0.33]{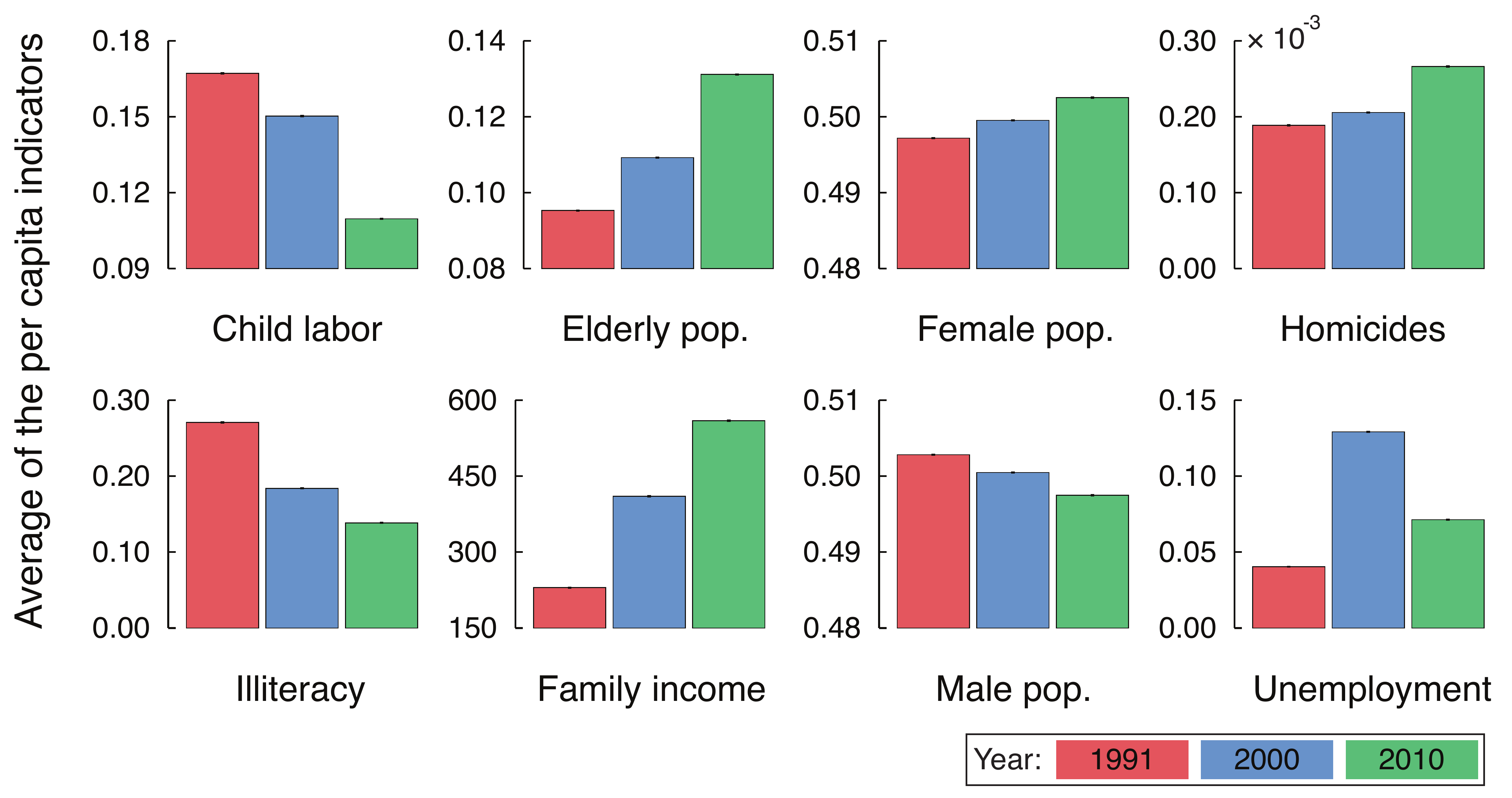}
\end{center}
\caption{
{\bf Trends in the evolution of per capita values of Brazilian urban indicators.} The bar plots show the temporal evolution of eight urban indicators in the years 1991, 2000 and 2010 in which the national census took place. The bar plots are average values over Brazilian cities, where each urban indicator for each city have been divided by the corresponding city population, that is, made per capita. The definition of each indicator is provided in the \hyperref[met:data]{Methods Section}. The tiny error bars are 95\% bootstrap confidence intervals for the average values. We note that elderly population, female population, homicides and family income display an increasing tendency; whereas child labor, illiteracy and male population have decreased over time. Despite having increased, the unemployment rates show a not very clear trend.
}
\label{fig:1}
\end{adjustwidth}
\end{figure}

We start by considering the average of the per capita values for eight urban indicators described in the \hyperref[met:data]{Methods Section}. This is a common practice of government agencies for tracking the evolution of a particular city or for comparing a group of cities with different populations. We observe in Fig.~\ref{fig:1} that almost all per capita indicators show a clear temporal trend: elderly population, female population, homicides and family income have increased over the years; whereas child labor, illiteracy and male population have decreased (the unemployment rates have evolved in a more complex manner, exhibiting no clear tendency). We could also list the cities in which these indicators have considerably changed or rank the cities that have made more progress in reducing, for instance, the homicide or illiteracy rates. 

One of the main problems with this analysis is that it completely ignores the hypothesis that most urban indicators $Y_i(t)$ displays allometric relationships (or scale invariance) with the population $N(t)$, that is, 
\begin{equation}
Y_i(t) = 10^{\mathcal{A}_i}\, N(t)^{\beta_i}\,,
\end{equation}
where $\mathcal{A}_i$ is a constant, $\beta_i$ is the allometric (or scaling) exponent and $t$ stands for time. This simple relationship summarizes the (average) effects of increasing the population size on the urban indicators; it states that cities are self-similar in terms of their population, in the sense that average properties of a given city can be inferred only by knowledge of its population. Urban indicators are thus expected to display a deterministic component emerging from very few and general properties of the urban networks related to social and infrastructure aspects of cities~\cite{Bettencourt4}. When analyzing per capita values, it is implicitly assumed that the value of an urban indicator is proportional to the population size ($\beta_i=1$) or, in other words, that cities are extensive systems. This idea is opposed to the complex systems approach of cities: complex systems are non-extensive ($\beta_i\neq1$), meaning that its isolated parts do not behave in the same manner as when they are interacting. Cities have similar properties and only make sense as an entire ``organism'' and, in fact, there is robust evidence favoring the non-extensive and universal nature of cities (that is, the urban scaling hypothesis) across different cultures and historical periods~\cite{Bettencourt4,Bettencourt5,Ortman}. Thus, several properties of a city of a given size cannot be linearly scaled for another city with larger or smaller population size. The dynamical processes mediated by the urban networks make the scale operation a nonlinear transformation for several urban indicators, often resulting in per capita savings of material infrastructure and in gain of socio-economic productivity~\cite{Bettencourt}. From a more technical point of view, whenever the allometric exponent $\beta_i$ is different from one, there is a remaining component related to the population size when evaluating the per capita values of these urban indicators, that is, $y_i=Y_i / N \sim N^{(\beta_i-1)}$, which creates a bias towards large cities for $\beta_i>1$ and towards small cities for $\beta_i<1$; per capita measures are only efficient in correctly removing the effect of the population size in an urban indicator for $\beta_i=1$. For our data, a complete description of the allometric relationships between the eight urban indicators and the population is presented in the \hyperref[met:plfit]{Methods Section}, where we have confirmed the presence of allometries in our data as summarized in Table~\ref{tab:allometric} (see also \hyperref[S1_Fig]{S1} and \hyperref[S2_Fig]{S2~Figs.}). 

\begin{table}[!ht]
\caption{
{\bf Allometric relationships between urban indicators and population size.} Values of parameters $\mathcal{A}_i$ and $\beta_i$ obtained via orthogonal distance regression on the relationship between $\log{Y_i(t)}$ and $\log{N(t)}$ for each urban indicator in the year $t$ (see 
\hyperref[met:plfit]{Methods Section}). The values inside the brackets are the standard errors (SE) in the last decimal of the estimated parameters. The last column shows the values of the Pearson linear correlation coefficient $\rho$ for each allometry in log-log scale.
}
\renewcommand{\arraystretch}{0.8}

\centering
\begin{tabular}{lcrrc}
\hline
Indicator $Y_i(t)$ & Year $t$ & $\mathcal{A}_i\,(\text{SE})$ & $\beta_i\,(\text{SE})$ & $\rho$ \\
\hline
\multirow{3}{*}{Child labor} & 1991 & $-0.64\,(5)$ & $0.96\,(1)$ & 0.909 \\
 & 2000 & $-0.55\,(5)$ & $0.93\,(1)$ & 0.906 \\
 & 2010 & $-0.80\,(5)$ & $0.95\,(1)$ & 0.884 \\
 \hline
\multirow{3}{*}{Elderly population} & 1991 & $-0.99\,(5)$ & $0.992\,(6)$ & 0.976 \\
 & 2000 & $-0.83\,(2)$ & $0.969\,(5)$ & 0.980 \\
 & 2010 & $-0.72\,(2)$ & $0.963\,(5)$ & 0.982 \\
 \hline
\multirow{3}{*}{Female population} & 1991 & $-0.367\,(3)$ & $1.014\,(1)$ & 1.000 \\
 & 2000 & $-0.361\,(2)$ & $1.013\,(1)$ & 1.000 \\
 & 2010 & $-0.355\,(2)$ & $1.012\,(1)$ & 1.000 \\
 \hline
\multirow{3}{*}{Homicides} & 1991 & $-5.4\,(1)$ & $1.35\,(3)$ & 0.769 \\
 & 2000 & $-5.7\,(1)$ & $1.41\,(2)$ & 0.800 \\
 & 2010 & $-5.04\,(9)$ & $1.29\,(2)$ & 0.827 \\
 \hline
\multirow{3}{*}{Illiteracy} & 1991 & $-0.29\,(7)$ & $0.92\,(2)$ & 0.789 \\
 & 2000 & $-0.26\,(7)$ & $0.87\,(2)$ & 0.774 \\
 & 2010 & $-0.28\,(7)$ & $0.85\,(2)$ & 0.749 \\
 \hline
\multirow{3}{*}{Family income} & 1991 & $0.82\,(6)$ & $0.33\,(1)$ & 0.428 \\
 & 2000 & $1.14\,(6)$ & $0.31\,(1)$ & 0.440 \\
 & 2010 & $1.63\,(5)$ & $0.23\,(1)$ & 0.440 \\
 \hline
\multirow{3}{*}{Male population} & 1991 & $-0.239\,(3)$ & $0.987\,(1)$ & 1.000 \\
 & 2000 & $-0.243\,(2)$ & $0.987\,(1)$ & 1.000 \\
 & 2010 & $-0.249\,(2)$ & $0.988\,(1)$ & 1.000 \\
 \hline
\multirow{3}{*}{Unemployment}  & 1991 & $-3.50\,(8)$ & $1.45\,(2)$ & 0.880 \\
 & 2000 & $-2.07\,(5)$ & $1.25\,(1)$ & 0.940 \\
 & 2010 & $-2.09\,(5)$ & $1.20\,(1)$ & 0.931 \\
 \hline
\end{tabular}
\label{tab:allometric}
\end{table}

\begin{figure}[!ht]
\begin{adjustwidth}{-2.25in}{0in}
\begin{center}
\includegraphics[scale=0.54]{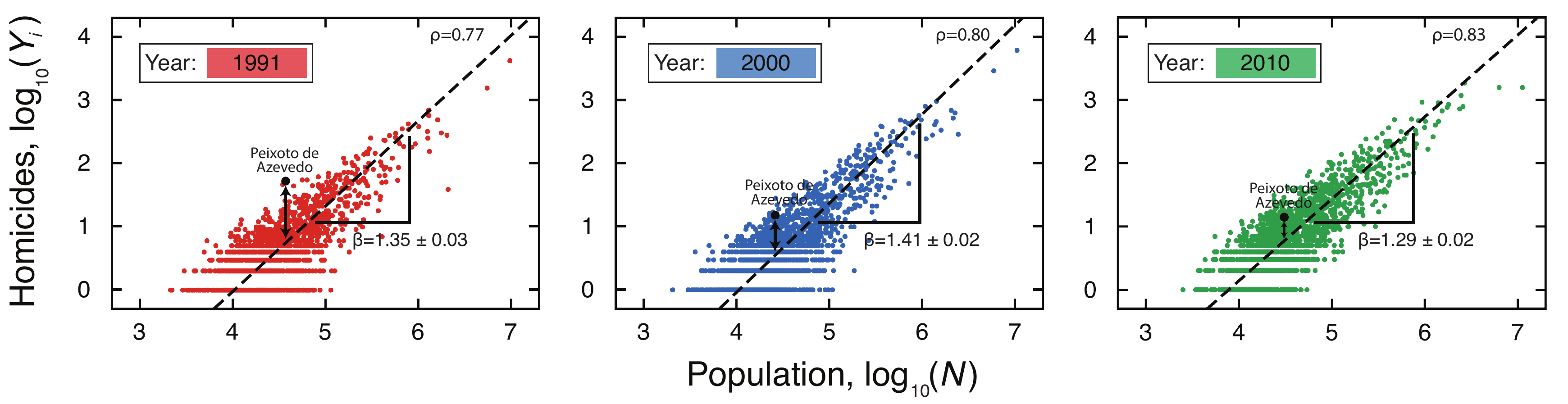}
\end{center}
\caption{
{\bf Allometric laws and the definition of the scale-adjusted metric $D_{Y_i}(t)$.} The scatter plots show the allometric relationships between number of homicides and population size for the years $t=1991$, $2000$ and $2010$ in log-log scale (see \hyperref[S1_Fig]{S1} and \hyperref[S2_Fig]{S2~Figs.} for all other indicators). The allometric exponents $\beta_i$ (see \hyperref[met:plfit]{Methods Section} for details on the calculation of $\beta_i$) and Pearson correlation coefficient $\rho$ are shown in the figures. We highlight a particular city (Peixoto de Azevedo) in the three years to illustrate the definition and the evolution $D_{Y_i}(t)$. For this city, the number homicides was quite above the allometric law in the year $t=1991$; however, it has approached the expected value by the allometric law over the years.
}
\label{fig:2}
\end{adjustwidth}
\end{figure}

The problem with per capita indicators thus prompts the question on how we can account for these allometries and correctly remove the effect of the population size on the urban indicators. Bettencourt \textit{et al.}~\cite{Bettencourt3} have proposed a simple and efficient procedure for overcoming this problem by defining the so-called \textit{scale-adjusted urban indicator}. The approach consists in evaluating the logarithmic difference between the actual value of an urban indicator $Y_i(t)$ and the value expected by allometric relationship with the population $N(t)$ (that is, the residuals in the allometric relationships) in given year $t$, namely,
\begin{equation}
D_{Y_i}(t)= \log Y_i(t)-[\mathcal{A}_i+\beta_i \, \log N(t)]\,.
\end{equation}
The previous quantity explicitly considers the allometry between an urban indicator and the population size, creating a relative measure that is not biased by the population size (size-independent) for any value of $\beta_i$. The scale-adjusted metric $D_{Y_i}(t)$ captures the exceptionality (either good or bad) of a city, which somehow is the result of the nonlinear agglomeration process related to the socio-economic choices and historical path of a city~\cite{Bettencourt3}. Furthermore, $D_{Y_i}$ establish a more ``natural" scale for ranking cities by identifying whether an urban indicator of a given city is above ($D_{Y_i}>0$) or below ($D_{Y_i}<0$) the expected value from cities of similar sizes. This approach already have proved to be useful in the economic context~\cite{Lobo,Podobnik} and was also employed for unveiling relationships between crime and urban metrics that are not properly carried out by linear regression analysis~\cite{Alves2}. Figure~\ref{fig:2} illustrates the definition of $D_{Y_i}(t)$ and shows an example of allometry between homicides and population size (see also \hyperref[S1_Fig]{S1} and \hyperref[S2_Fig]{S2~Figs.} for all urban indicators). 


\begin{figure}[!ht]
\begin{adjustwidth}{-2.25in}{0in}
\begin{center}
\includegraphics[scale=0.33]{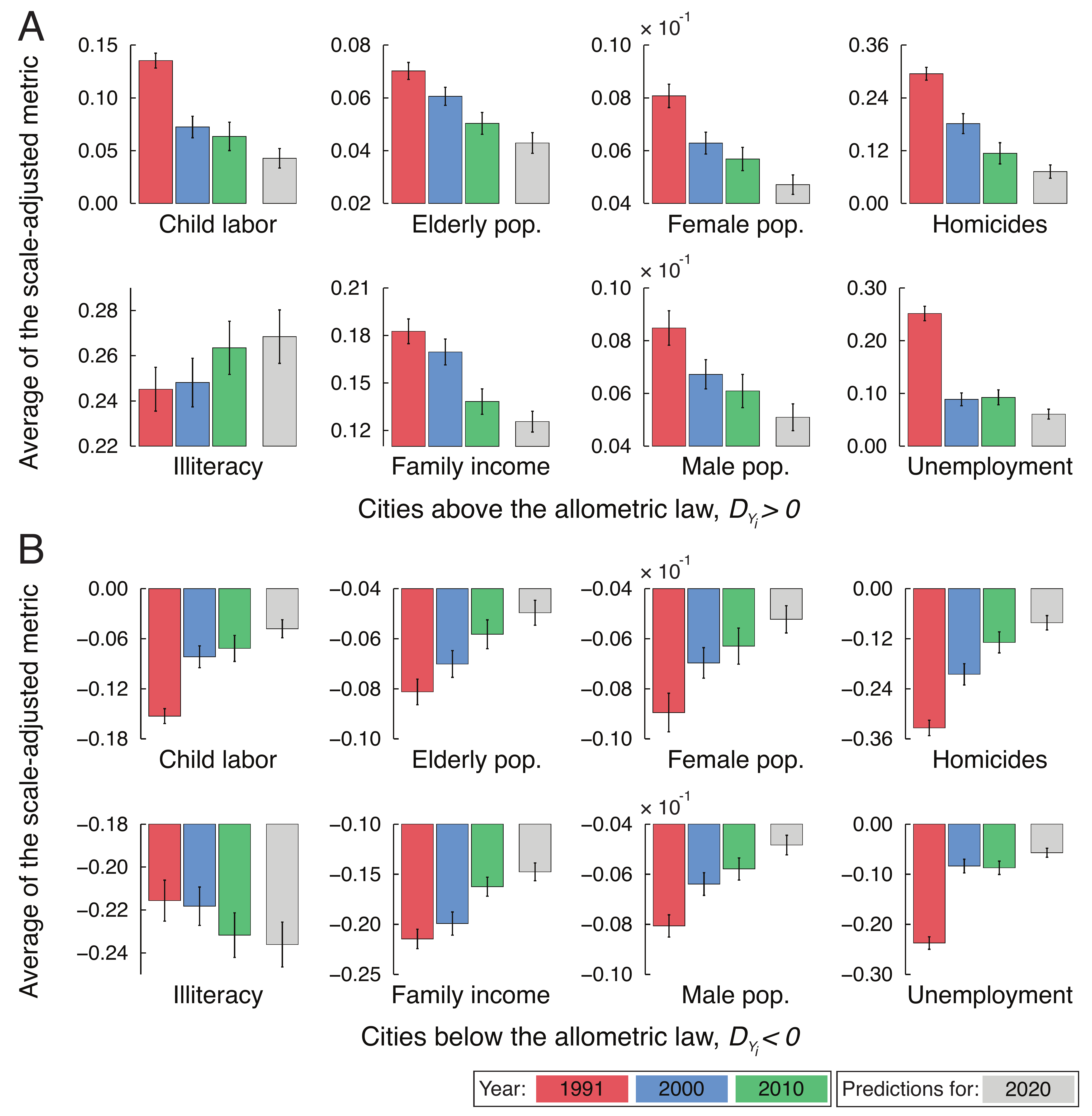}
\end{center}
\caption{
{\bf Trends in the evolution of the average values of the scale-adjusted metrics.} The bar plots show the evolution of the average $D_{Y_i}(t)$ when grouping the cities whose urban indicator was (A) above and (B) below the allometric law in the year $t=1991$. The colorful bars are empirical data (red for $t=1991$, blue for $t=2000$ and green for $t=2010$) and gray bars represent the predictions obtained from the linear model of Eq.~\ref{eq:glinmodel} (see discussion in main text). The error bars are 95\% bootstrap confidence intervals for the average values. Note that the average  $D_{Y_i}(t)$ displays a statistically significant decreasing tendency for cities whose urban indicator was initially above the allometric law and an increasing tendency is observed for cities whose urban indicator was initially below the allometric law. The only exception to this pattern is the case of illiteracy, where the average $D_{Y_i}(t)$ has increased for cities initially above the allometric law and it is almost a constant for those cities initially below. We further observe that the predicted values (gray bars) keep this main tendency.
}
\label{fig:3}
\end{adjustwidth}
\end{figure}

In order to show the informations that this scale-adjusted metric provides, we study the evolution of the average $D_{Y_i}(t)$ by creating two groups of cities: those whose the urban indicator $Y_i(t)$ was above ($D_{Y_i}(t)>0$) and those whose $Y_i(t)$ was below ($D_{Y_i}(t)<0$) the allometric law in the year $t=1991$. Figure~\ref{fig:3} shows these averages for the eight urban indicators over the three years in our database. For the majority of the urban indicators, the average $D_{Y_i}(t)$ displays a statistically significant decreasing tendency for cities initially above the allometric law; whereas an increasing tendency for the average $D_{Y_i}(t)$ is observed for those cities that were initially below the allometric law. For instance, cities where the number of children laboring, homicides and unemployment were above the value expected by the allometric law have been successful (on average) in reducing them; in contrast, cities where these indicators were below the allometric laws proved unable (on average) to improve or maintain this situation. The case of illiteracy is an exception to the previous pattern, since the average $D_{Y_i}(t)$ has increased for cities initially above the allometric law and it is almost a constant for those cities initially below. Thus, cities where illiteracy was initially above the allometric law have failed (on average) in increasing the number of literates; on the other hand, those cities initially below the allometric law have not only kept this feature (on average) but also managed to further improve the literacy levels. In the case of population metrics, particularly regarding female and male populations, the approaching to the allometric laws (together with the decrease in proportion of males in the population) may represent a good aspect with respect to reduction of violence, since an excessive contingent of men can drive an increase in  antisocial behavior~\cite{Hesketh}; moreover, both male population above and female population below the allometric law correlate with number of homicides above the allometric law~\cite{Alves2}. Similarly, family income above the allometric law correlates with homicides above, while family income below the allometric law correlates with homicides below~\cite{Alves2}. Remarkably, these informations remain hidden or distorted when we look only at the per capita values of these urban indicators.

\begin{figure}[!h]
\begin{adjustwidth}{-2.25in}{0in}
\begin{center}
\includegraphics[scale=0.4]{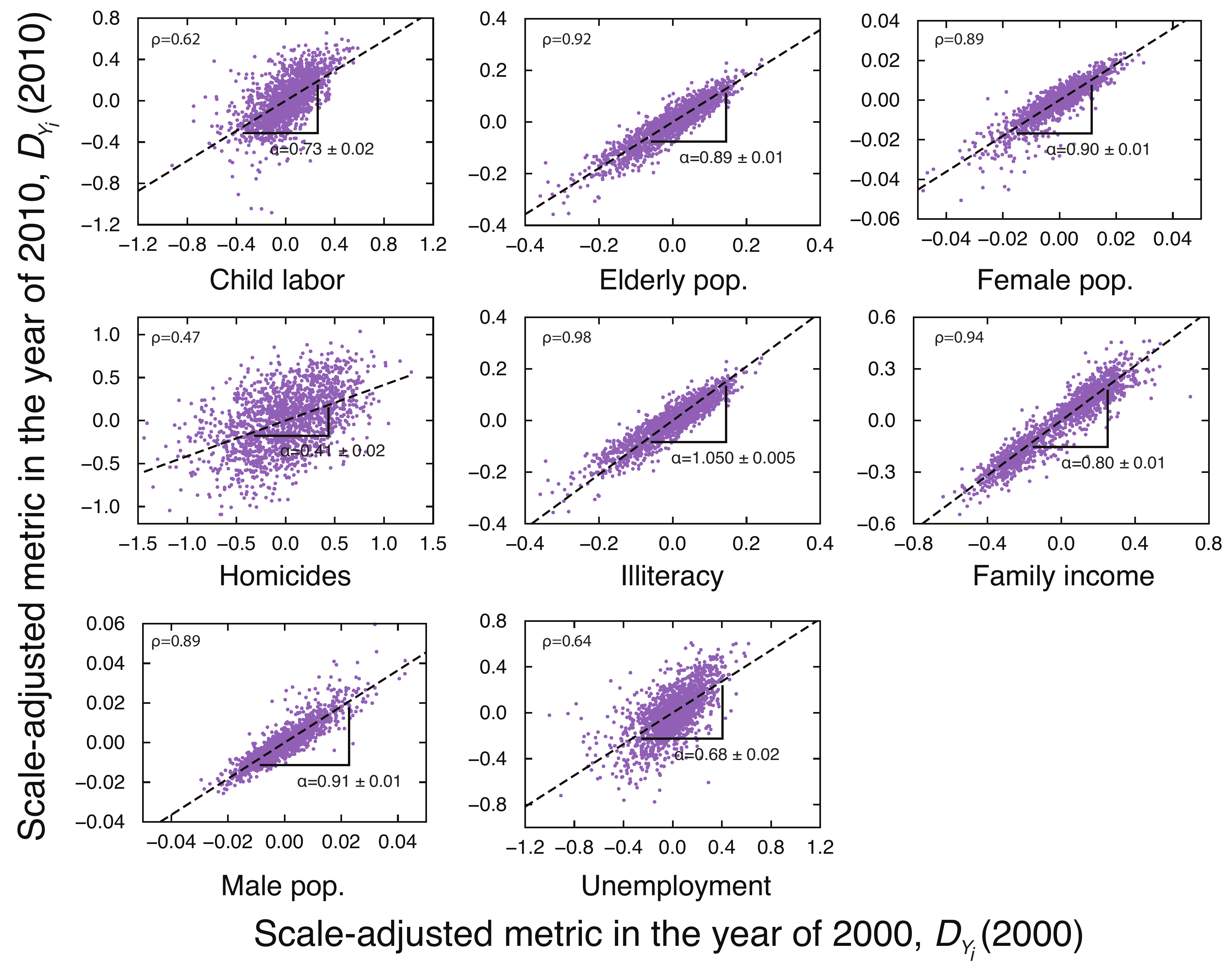}
\end{center}
\caption{
{\bf Memory effects in the evolution of the scale-adjusted metric $D_{Y_i}(t)$.} The purple dots show the values of $D_{Y_i}(2010)$ versus $D_{Y_i}(2000)$ for each city (see \hyperref[S3_Fig]{S3~Fig.} for $D_{Y_i}(2000)$ versus $D_{Y_i}(1991)$). Despite the different scattering degrees (see the value of Pearson correlation coefficient $\rho$ in the plots), we observe a linear correspondence between $D_{Y_i}(2010)$ and $D_{Y_i}(2000)$. The dashed lines are fits of the linear model \mbox{$D_{Y_i}(2010) = A_i + \alpha_i \, D_{Y_i}(2000)$} (Eq.~\ref{eq:linmodeldistance}) obtained via ordinary least-square regression. The values of $\alpha_i$ and their standard errors are shown in the plots and also summarized in Table~\ref{tab:2}. We note that $\alpha_i<1$ for almost all indicators, a fact that is in agreement with the approach to allometric laws observed for the averages value of $D_{Y_i}(t)$ (Fig.~\ref{fig:3}). Illiteracy is the only exception where $\alpha_i\gtrsim1$. This result also agrees with the departing from the allometric law of the average $D_{Y_i}(t)$ in the case of illiteracy.
}
\label{fig:4}
\end{adjustwidth}
\end{figure}

The trends in the average values of the scale-adjusted metrics prompt the question of how the values of this relative metric are affected by their previous values, that is, are the values of $D_{Y_i}(t+\Delta t)$ and $D_{Y_i}(t)$ correlated in some particular fashion? To answer this question, we analyze the scatter plots of the scale-adjusted metric in a given year versus its past values for each urban indicator for all cities. Figure~\ref{fig:4} shows these scatter plots considering the values of $D_{Y_i}(2010)$ versus $D_{Y_i}(2000)$ and \hyperref[S3_Fig]{S3~Fig.} shows the plots for $D_{Y_i}(2000)$ versus $D_{Y_i}(1991)$. We observe that, despite the different scattering degrees  (Pearson correlation ranging from $0.47$ to $0.98$), linear functions are good approximations for the average tendency of these relationships. We thus adjust the linear model
\begin{equation}\label{eq:linmodeldistance}
D_{Y_i}(t+\Delta t) = A_i + \alpha_i \, D_{Y_i}(t)\,
\end{equation}
to each urban indicator via least square method and the best fitting parameter $\alpha_i$ (Pearson correlations as well) is shown in Table~\ref{tab:2} for the two combinations of years. We have omitted the values of $A_i$ because they are very small ($\sim 10^{-6}$). Despite the increasing tendency observed for the values of $\alpha_i$ over the years (expect for homicides and family income), we observe that $\alpha_i<1$ for almost all urban indicators except for illiteracy ($\alpha_i=1.01\pm0.01$ for 2000 versus 1991 and $\alpha_i=1.05\pm0.01$ for 2010 versus 2000). These results agree with the evolution of the average $D_{Y_i}(t)$, that is, indicators characterized by $\alpha_i<1$ present a tendency of approaching the allometric laws, while for $\alpha_i\geq1$ there should be a departing tendency from the allometric laws. 

\begin{table}
\renewcommand{\arraystretch}{.8}
\caption{{\bf Linear coefficients $\alpha_i$ of the memory relationships between the scale-adjusted metrics in different years.} Values of the parameters $\alpha_i$ obtained via least square fitting the model of the Eq.~\ref{eq:linmodeldistance} for the relationships between the scale-adjusted metrics: $D_{Y_i}(2000)$ versus $D_{Y_i}(1991)$ and $D_{Y_i}(2010)$ versus $D_{Y_i}(2000)$. We have omitted the values of the parameters $A_i$ because they are very small ($\sim 10^{-6}$). The values inside the brackets are the standard errors (SE) in the last decimal of the estimated parameter. The last column shows the values of the Pearson linear correlation coefficients $\rho$ for each relationship.
}

\centering
\begin{tabular}{lcrr}
\hline
{Indicator} & {Years} & $\alpha_i\,(\text{SE})$  & $\rho$ \\
\hline
\multirow{2}{*}{Child labor} & {$2000$\,--\,$1991$} & 0.54\,(2) & 0.53 \\
 & {$2010$\,--\,$2000$} & 0.73\,(2) & 0.62 \\
\hline
\multirow{2}{*}{Elderly population} & {$2000$\,--\,$1991$} & 0.84\,(1) & 0.92\\
 & {$2010$\,--\,$2000$} & 0.89\,(1) & 0.92\\
\hline
\multirow{2}{*}{Female population} & {$2000$\,--\,$1991$} & 0.66\,(1) & 0.83 \\
 & {$2010$\,--\,$2000$} & 0.90\,(1) & 0.89 \\
\hline
\multirow{2}{*}{Homicides} & {$2000$\,--\,$1991$} & 0.60\,(2) & 0.59 \\
 & {$2010$\,--\,$2000$} & 0.41\,(2) & 0.47 \\
\hline
\multirow{2}{*}{Illiteracy} & {$2000$\,--\,$1991$} & 1.010\,(5) & 0.98 \\
 & {$2010$\,--\,$2000$} & 1.050\,(5) & 0.98 \\
\hline
\multirow{2}{*}{Family income} & {$2000$\,--\,$1991$} & 0.89\,(1) & 0.91\\
 & {$2010$\,--\,$2000$} & 0.80\,(1) & 0.94 \\
\hline
\multirow{2}{*}{Male population} & {$2000$\,--\,$1991$} & 0.70\,(1) & 0.84 \\
 & {$2010$\,--\,$2000$} & 0.91\,(1) & 0.89 \\
\hline
\multirow{2}{*}{Unemployment} & {$2000$\,--\,$1991$} & 0.34\,(1) & 0.53 \\
 & {$2010$\,--\,$2000$} & 0.68\,(2) & 0.64 \\
\hline
\end{tabular}
\label{tab:2}
\end{table}

In order to better understand the role of the parameter $\alpha_i$, we consider the limit where $\Delta t$ is small to rewrite Eq.~\ref{eq:linmodeldistance} as the following differential equation
\begin{equation}\label{eq:difdistance}
\frac{d}{dt} D_{Y_i}(t) = A_i + (\alpha_i-1)D_{Y_i}(t)\,,
\end{equation}
whose solution is
\begin{eqnarray}\label{eq:soldistance}
D_{Y_i}(t) = 
\begin{cases}
A_i/(1-\alpha_i)+\left[k-A_i/(1-\alpha_i)\right]\,\exp[{-(1-\alpha_i)t}]~~~(\text{for}~\alpha \neq 1)\,, \\
A_i \,t + k~~~(\text{for}~\alpha = 1)\,,
\end{cases}
\end{eqnarray}
where $k$ is an integration constant. Thus, Eq.~\ref{eq:soldistance} predicts that $D_{Y_i}(t)$ will exponentially approach the value $A_i/(1-\alpha_i)$ for $\alpha_i<1$ and that they will exponentially increase over time when $\alpha_i>1$. For $\alpha_i=1$ we have a linear behavior for the evolution of $D_{Y_i}(t)$. It is worth remembering that $A_i$ is a very small number ($\sim10^{-6}$) for all urban indicators and hence the values of $D_{Y_i}(t)$ are actually approaching zero for $\alpha_i<1$ (that is, the values of the urban indicators are getting closer to value expected by the allometric law). We further observe that $1/(1-\alpha_i)$ plays the role of a characteristic time and if we assume $\alpha_i<1$, the smaller the value of $\alpha_i$ is, the faster we expect to be changes in the urban indicator. For population-related indicators, we observe an increasing tendency in the values of $\alpha_i$ and also that these values are among the largest ones, which agrees with the reduction of the Brazilian population growth in the last decades. For socio-economic indicators, we have, on one hand, that the values of $\alpha_i$ for child labor and unemployment have also increased (pointing out to slowdown of their dynamics); on the other hand, homicides and family income had their values of $\alpha_i$ reduced, suggesting that their rates of change have increased. Apart from the evolution in the values of $\alpha_i$, population-related indicators have (in general) larger values of $\alpha_i$ than those observed for socio-economic indicators, indicating that these latter are more susceptible to natural or public policies driven changes. The illiteracy is unique because its value of $\alpha_i$, that was very close to one (for $D_{Y_i}(2000)$ versus $D_{Y_i}(1991)$), has increased, a result that agrees with the detachment from the allometry reported in Fig.~\ref{fig:3} and suggests an acceleration in this process.

In addition of being autocorrelated with their past values, $D_{Y_i}(t+\Delta t)$ for the urban indicator $i$ also displays statistically significant crosscorrelations with $D_{Y_j}(t)$ for other indicators $j$ (\hyperref[S4_Fig]{S4~Fig.}). These memory effects and also the fact that the residuals surrounding the relationships $D_{Y_i}(t+\Delta t)$ versus $D_{Y_i}(t)$ are very close to Gaussian distributions (\hyperref[S5_Fig]{S5~Fig.}) with standard deviations across windows practically constant (\hyperref[S6_Fig]{S6~Fig.}) make these scale-adjusted metrics particularly good for being used in linear regressions aiming forecasts. We have thus adjusted the linear model (via ordinary least-squares method)
\begin{equation}\label{eq:glinmodel}
D_{Y_i}(t+\Delta t) = C_0 + \sum_{k=1}^8 C_k\, D_{Y_k}(t) + \eta_i(t)\,
\end{equation}
by considering the relationships $(t+\Delta t)=2000$ versus $t=1991$ and $(t+\Delta t)=2010$ versus $t=2000$. In Eq.~\ref{eq:glinmodel}, $C_k$ is the linear coefficient quantifying the predictive power of $D_{Y_k}(t)$ on $D_{Y_i}(t+\Delta t)$ ($C_0$ is the intercept coefficient) and $\eta_i(t)$ is the noise term accounting for the effect of unmeasurable factors. The results exhibiting the linear coefficients of each linear regression for the two combinations of years are shown in \hyperref[S1_Text]{S1 Text}. We note that these simple models account for 31\%--97\% of the observed variance in $D_{Y_i}(t+\Delta t)$ and that they correctly reproduce the average values of the scale-adjusted metric above and below the allometric laws for the years 2000 and 2010 only using data from the years 1991 and 2000, respectively (see \hyperref[S7_Fig]{S7~Fig.}). We have further compared the distributions of the empirical values of $D_{Y_i}$ with the predictions of these linear models and observed that the agreement is remarkable good for the indicators elderly, female and male population as well as for illiteracy and income (\hyperref[S8_Fig]{S8} and \hyperref[S9_Fig]{S9~Figs.}). Motivated by these good agreements, we proposed to forecast the values of $D_{Y_i}(t+\Delta t)$ in the year of 2020 (next Brazilian national census). In order to do so, we have considered that the linear coefficients $C_k$ are constant over time and employed the average value of $C_k$ over the two combinations of years used in Eq.~\ref{eq:glinmodel} for predicting the values of $D_{Y_i}(t+\Delta t)$. It is worth noting that by assuming $C_k$ constant, we are ignoring the evolution of socio-economic and policy factors. In an ideal scenario, one could track the evolution of the values of $C_k$ for achieving more reliable predictions. However, our data (that is, the two values for $C_k$) do not enable us to probe possible evolutionary behaviors in the values of $C_k$. Even so, as pointed by Bettencourt~\cite{Bettencourt3}, the dynamics of the urban metrics seems to be dominated by long timescales ($\sim$30 years), and thus the approach of constant coefficients should be seen as a first approximation. The grays bars in Fig.~\ref{fig:3} show the averages $D_{Y_i}(2020)$ after grouping the cities with $D_{Y_i}(1991)>0$ and $D_{Y_i}(1991)<0$. We observe that predictions for the average values basically keep the trends presented in the previous years; for unemployment, in which the trend was not very clear, the predictions put this indicator together with most indicators, where the average $D_{Y_i}(t)$ has been decreasing for cities initially above the allometric law and increasing for those initially below. 

\begin{figure}[!ht]
\begin{adjustwidth}{-2.25in}{0in}
\begin{center}
\includegraphics[scale=0.4]{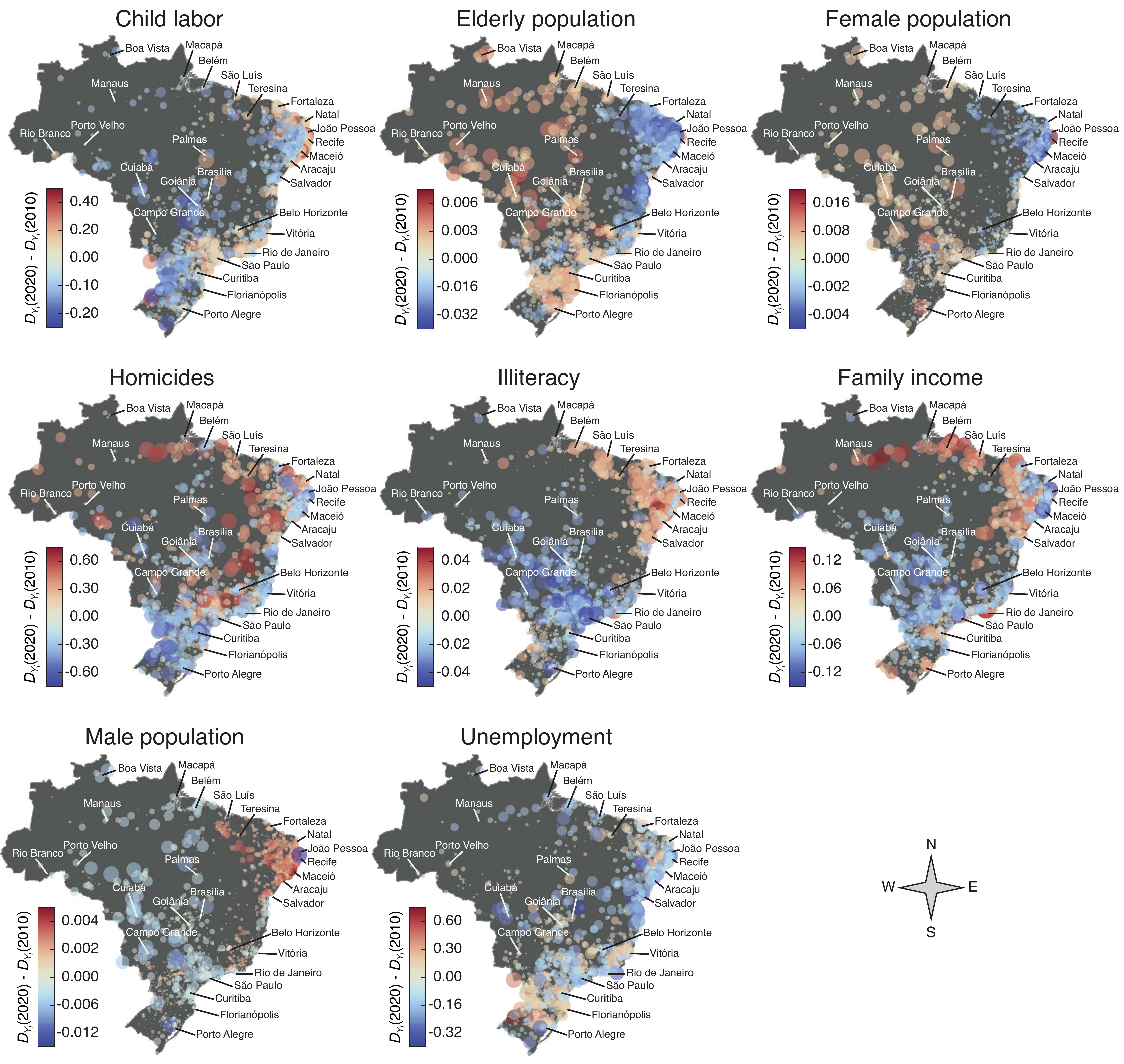}
\end{center}
\caption{
{\bf Geographic visualization of the predicted changes in the scale-adjusted metrics $D_{Y_i}(t)$ between the years $t=2010$ and $t=2020$.} Each circle represents a city and the radius of the circle is proportional to $|D_{Y_i}(2020)-D_{Y_i}(2010)|$. We color the circles according to the difference between $D_{Y_i}(2020)$ and $D_{Y_i}(2010)$: shades of azure indicate that we expect a decrease in the values of $D_{Y_i}(2020)$, whereas shades of red show the cities where we expect an increase in the values of $D_{Y_i}(2020)$. The labeled cities are the capitals of the twenty-seven Federal Units of Brazil (the Brazilian states and the federal district). The forecast for the values of $D_{Y_i}(2020)$ were obtained through the linear model of Eq.~\ref{eq:glinmodel}, where the linear coefficient $C_k$ were averaged over the two combinations of years $2000$\,--\,$1991$ and $2010$\,--\,$2000$. We note that the changes in $D_{Y_i}(t)$ appear spatially clustered for most indicators, forming regions where most cities are expected to increase or decrease the value of the scale-adjusted metric.
}
\label{fig:5}
\end{adjustwidth}
\end{figure}

In order to gain further information on the predictions, we build a geographic visualization of the expected changes in $D_{Y_i}(t)$ between the years  $t=2010$ and $t=2020$. The circles over the maps in Fig.~\ref{fig:5} show the geographic location of Brazilian cities; the radii of these circles are proportional to $|D_{Y_i}(2020)-D_{Y_i}(2010)|$ and are colored with shades of azure for cities where $[D_{Y_i}(2020)-D_{Y_i}(2010)]<0$ (the indicator is expected to decrease) and with shades of red for cities where $[D_{Y_i}(2020)-D_{Y_i}(2010)]>0$ (the indicator is expected to increase); in both cases, the darker the shade, the larger is the absolute value of the difference $[D_{Y_i}(2020)-D_{Y_i}(2010)]$. Perhaps, the most striking feature of these visualizations is the fact that the predicted changes appear spatially clustered for almost all indicators, which somehow reflects the geographic inequalities existing in Brazil; however, some intriguing patterns are indicator-dependent. 

For child labour, $D_{Y_i}(t)$ is expected to increase around three of the most densely populated regions that contain the metropolitan areas of S\~ao Paulo, Rio de Janeiro and the metropolitan areas of almost all northeast capitals; we further observe that a decrease in child labour cases is expected in mostly of the inner and southern cities. For the indicators elderly, female and male populations the clustering of the changes in $D_{Y_i}(t)$ is quite evident:  elderly and female populations are foreseen to decrease in mostly of the northeast cities and display an increasing tendency in large part of the other regions (male population behaves anti-symmetrically to the female population). In the case of homicides, our model predicts a decrease in $D_{Y_i}(t)$ for the vast majority of southern cities and we further observe a stripe near the east coast where $D_{Y_i}(t)$ is expected to decrease for mostly of cities (both densely populated areas); on the other hand, inner cities (specially inner cities from the state of S\~ao Paulo and northeastern region) are expected to increase $D_{Y_i}(t)$, suggesting that this violent crime may be ``moving'' towards less populated areas of the interior of Brazil. 

For illiteracy, again, the clustering of the changes is easily perceptible: we expect an increase in $D_{Y_i}(t)$ for mostly of the northeast and northernmost cities, while a decrease is predicted for the majority of the cities from other regions (excluding several inner cities of southernmost region). For family income, we also observe a clustering in the changes where most northern cities are expected to increase the value of $D_{Y_i}(t)$ (specially the inner cities of these regions), while for most cites in the central part of Brazil are expected to decrease the value of $D_{Y_i}(t)$; these expected changes may be (at least in part) related to the ``bolsa fam\'ilia'' (family allowance) program --- a large scale social welfare program of the Brazilian government (more than 14 million families were beneficiaries in 2013) for providing financial aid to poor families via direct cash transfer --- because large part of the families receiving this aid are from the north and northeast regions. It is worth noting that for participating in the program, families must insure that their children attend to school and thus one would expect a reduction in $D_{Y_i}(t)$ for illiteracy in same regions that concentrate the beneficiaries of the program, which was not predicted by the model. This result suggest that simply enforce school attendance may not be efficient for reducing illiteracy, which can also be explained by common observed poor conditions of the public school system of those regions. Finally, for unemployment, there is no clear clustering of the changes in $D_{Y_i}(t)$, but instead we expect a decrease in $D_{Y_i}(t)$ widespread throughout the Brazilian territory (except in the southernmost region, where we note the prevalence of light shades of red).

\subsection*{Summary}
Our article discusses that despite being widely used in government reports and also in academic works, per capita indicators completely ignore the universal allometric laws that appear ruling the growth dynamics of cities. We discuss that per capita indicators can be biased towards small or large cities depending on whether we have sub or superlinear allometries with the population size. We thus employed a scale-adjusted metric by evaluating the difference between the actual values of urban indicators and the ones that are expected by the allometric relationships. When investigating the evolution of the scale-adjusted metrics, we have reported patterns that do not appear in the per capita indicators. The scale-adjusted metrics also display a linear correspondence with their past values, a feature that facilitates the use of linear regressions for modeling the urban indicators. By employing simple linear models for describing the scale-adjusted metrics based on their past values, we verified that these models account for 31\%--97\% of the observed variance and correctly reproduce the average values of the scale-adjusted metric. Assuming the linear coefficients constant over time, we present predictions for the values of the scale-adjusted metrics in year of 2020, when the next Brazilian census will happen. We observe that the predicted changes for the urban indicators appear (for most cases) spatially clustered, that is, forming regions where most cities are expected to increase or decrease the value of the scale-adjusted metric. Apart from this visualization of the predicted changes, we also provide a table (\hyperref[S1_dataset]{S1 Dataset}) with the values of the scale-adjusted metrics for the three past census data as well as the predictions for the next one in the supplementary materials. We believe that our analysis may find potential applications on development of new policies and resources allocation in the context of urban planing. Finally, we note that the methods worked out here can also be directly applied in other contexts where allometries are present such as for economic indexes and biological quantities. 

\section*{Methods}
\subsection*{Data presentation}\label{met:data}
The data we analyzed consist of the population size $N(t)$ and eight urban indicators $Y_i(t)$ for each Brazilian city in the years $t=1991$, $2000$ and $2010$ in which the national census took place. We filter these data by selecting the 1605 cities for which all the eight urban indicators were available, this corresponds to 28.8\% of the total number of Brazilian cities but account for 76.5\% of the total population of Brazil. These data are maintained and made freely available by the Department of Informatics of the Brazilian Public Health System --- DATASUS~\cite{datasus}. 
The eight urban metrics are defined as follows:  
\textit{Child labour:} the proportion of the population aged 10 to 15 years who is working or looking for work during the reference week, in a given geographic area, in the current year; \textit{Elderly population:} the number of inhabitant of a given city aged 60 years or older; \textit{Female population:} the number of inhabitant of a given city that is female; \textit{Homicides:} injuries inflicted by another person with intent to injure or kill, by any means. \textit{Illiteracy:} it gives the number of inhabitants in a given geographic area, in the current year, aged 15 years or older, who cannot read and write at least a single ticket in the language they know;  \textit{Family income:} this indicator gives the average household incomes of residents in a given geographic area, in the current year. It was considered as family income the sum of the monthly income of the household, in Reals (Brazilian currency) divided by the number of its residents; \textit{Male population:} the number of inhabitant of a given city that is male; \textit{Unemployment:} it gives the number of inhabitant aged 16 years or older who is without working or looking for work during the reference week, in a given geographic area, in the current year. 

Despite there being other definitions~\cite{Angel}, the results presented here have been obtained by considering that cities are the smallest administrative units with a local government (municipalities or \textit{munic\'ipios}). The other commonly employed definition is the metropolitan area, which is composed of more than one municipality and its is usually associated with the coalescence of several municipalities. As discussed by Bettencourt \textit{et al.}~\cite{Bettencourt5}, the choice of the ``unit of analysis'' is crucial when studying properties of cities. Regarding the scaling analysis: on one hand, the disaggregation of the correct urban definition can introduce a bias in the value of the scaling exponent (either by reducing or increasing its expectation value); on the other hand, the aggregation of the correct urban definition usually make the allometry more linear~\cite{Bettencourt5}. In fact, changes in the scaling exponents have been reported when choosing different definitions of city~\cite{Schlapfer,Louf3,Arcaute}. However, there is no fail-safe procedure for defining the correct boundaries of a city, also some urban indicators are actually more spatially restricted than others (for instance, homicides versus family income). We have also analyzed our data after considering simultaneously the municipalities that do not belong to any metropolitan area and by aggregating the municipalities of the 39 metropolitan areas existing in Brazil. Despite the observation of relatively small changes in the scaling exponents, our conclusions remain unaltered under this scenario. 


\subsection*{Fitting allometric laws between urban indicators and population}\label{met:plfit}
As we previously mentioned, several works have reported the existence of robust allometric relationships between several urban indicators and the population size. Regarding Brazilian cities, scaling laws already have been identified for several urban indicators~\cite{Bettencourt2,Gomez-Lievano,Alves,Alves2,Alves3,Mantovani,Mantovani2,Melo,Ignazzi}, mainly because of the existence of reliable data made freely available by Brazilian agencies (DATASUS and IBGE). Here, we want to confirm that the urban scaling hypothesis holds for all our urban indicators and if these allometries have been changed over time. Specifically, we test the hypothesis that an urban indicator can be described by a power-law function of the population size, that is, $Y_i(t) = 10^{\mathcal{A}_i}\, N(t)^{\beta_i}$, where $\beta_i$ is the allometric (or scaling) exponent and $\mathcal{A}_i$ is constant. In order to do so, we have plotted the logarithm of each urban indicator against the logarithm of the population and adjusted a linear model via orthogonal distance regression (as implement in the package \texttt{scipy.odr} of the Python library SciPy~\cite{SciPy}) to all these relationships. Although the empirical relationships present different scattering degrees, they all display good quality linear relationships (Pearson correlation ranging from $0.44$ to $1.00$ --- see Table~\ref{tab:allometric}) which are well described by linear models in log-log scale (see Fig~\ref{fig:2} and \hyperref[S1_Fig]{S1} and \hyperref[S2_Fig]{S2~Figs.}). From Table~\ref{tab:allometric}, we further note that the values of $\beta_i$ for illiteracy, family income and unemployment display a weak decreasing tendency over the years, while the other indicators show only small fluctuations (no clear evolutive tendency). A weak decreasing tendency for unemployment also appears in the work of Ignazzi~\cite{Ignazzi} on the same data from the years of 2000 and 2010. The values of $\beta_i$ thus classify our indicators in two groups: female population, homicides and unemployment have superlinear relationships with the population ($\beta_i>0$); while child labor, elderly population, illiteracy, family income and male population have sublinear ones ($\beta_i<0$). It is worth noting that despite the allometric exponents $\beta_i$ be close to one for elderly, female and male populations, the allometries between these indicators and total population are almost perfectly correlated, producing values of $\beta_i$ very close but statistically different from one. We further observe that the values of the allometric exponents $\beta_i$ reported here may slightly differ from previous-reported one due the different fitting procedures as well as different urban definitions; however, these discrepancies are often very small (for instance, by considering generalized least squares via the Cochrane-Orcutt procedure and another definition of city, Ignazzi~\cite{Ignazzi} have found $\beta=1.23$ and $\beta=1.19$ for unemployment, respectively in years of 2000 and 2010).


\section*{Authors Contributions}
H.V.R. and L.G.A.A. designed the research, analyzed the data and prepared the figures. All authors wrote and reviewed the manuscript.


%
%
%

\clearpage
\section*{Supporting Information}

\subsection*{S1 Dataset}
\label{S1_dataset}
{\bf Scale-adjusted metrics for Brazilian cities.} Values of the scale-adjusted metrics ($D_{Y_i}$) for the eight urban indicators of Brazilian cities in years of 1991, 2000 and 2010 as well as the predictions obtained via the linear model (Eq.~\ref{eq:glinmodel}) for year of 2020. 
\clearpage

\begin{adjustwidth}{-2.25in}{0in}
\subsection*{S1 Fig.}
\label{S1_Fig}
\begin{center}
\includegraphics[scale=0.45]{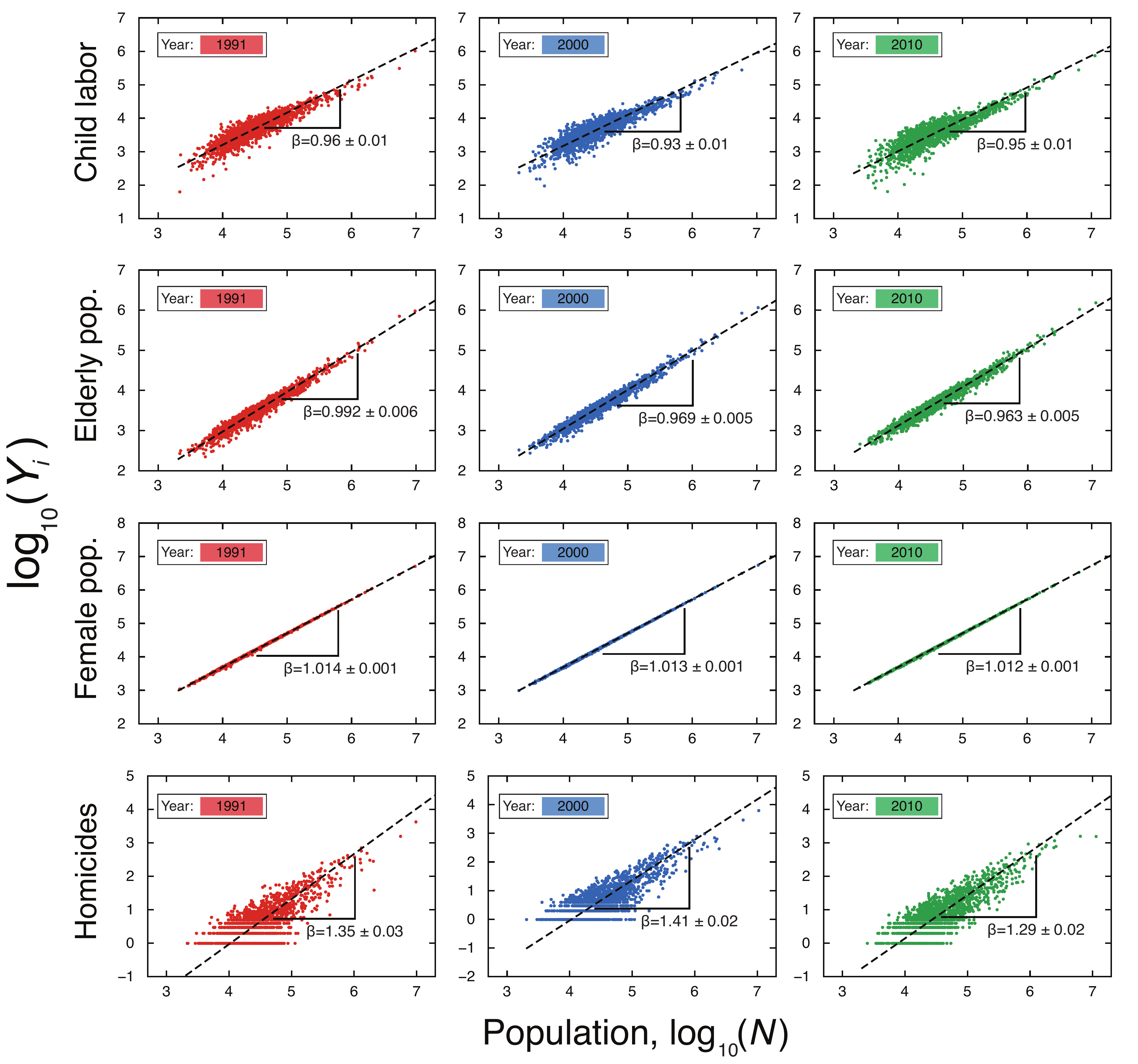}
\end{center}

\noindent {\bf Allometric laws with the population size.} The scatter plots show the allometric relationships between the urban indicators (from top to bottom: child labor, elderly population, female population and homicides) and population size for the years $t=1991$ (red dots), $2000$ (blue dots) and $2010$ (green dots) in log-log scale. The allometric exponents $\beta_i$ (see \hyperref[met:plfit]{Methods Section} for details on the calculation of $\beta_i$) are shown in the figures. See \hyperref[S2_Fig]{S2 Fig.} for the other indicators.
\end{adjustwidth}
\clearpage

\begin{adjustwidth}{-2.25in}{0in}
\subsection*{S2 Fig.}
\label{S2_Fig}
\begin{center}
\includegraphics[scale=0.45]{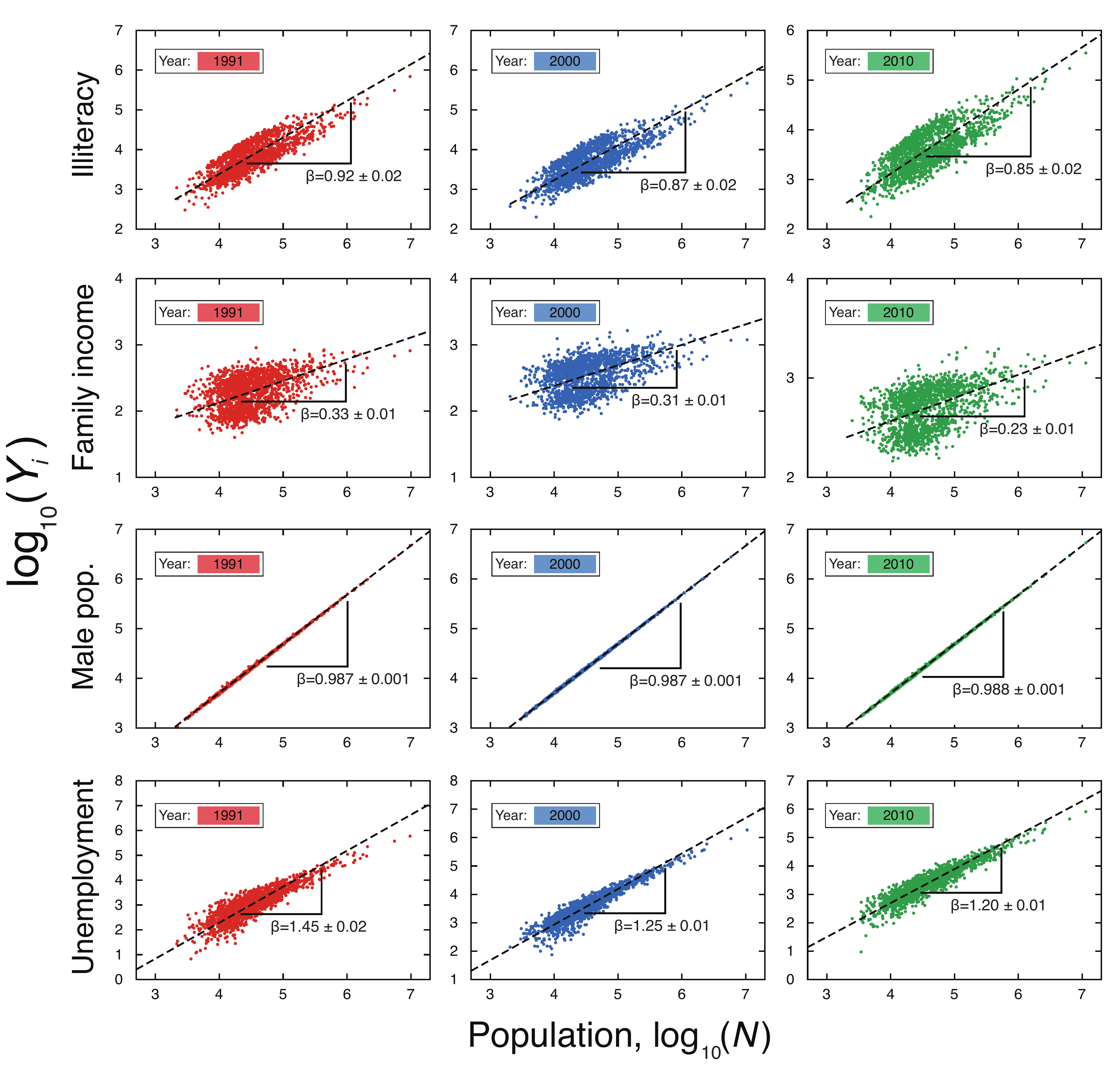}
\end{center}

\noindent {\bf Allometric laws with the population size.} The same as \hyperref[S1_Fig]{S1 Fig.} for the indicators illiteracy, family income, male population and unemployment.
\end{adjustwidth}
\clearpage

\begin{adjustwidth}{-2.25in}{0in}
\subsection*{S3 Fig.}
\label{S3_Fig}
\begin{center}
\includegraphics[scale=0.5]{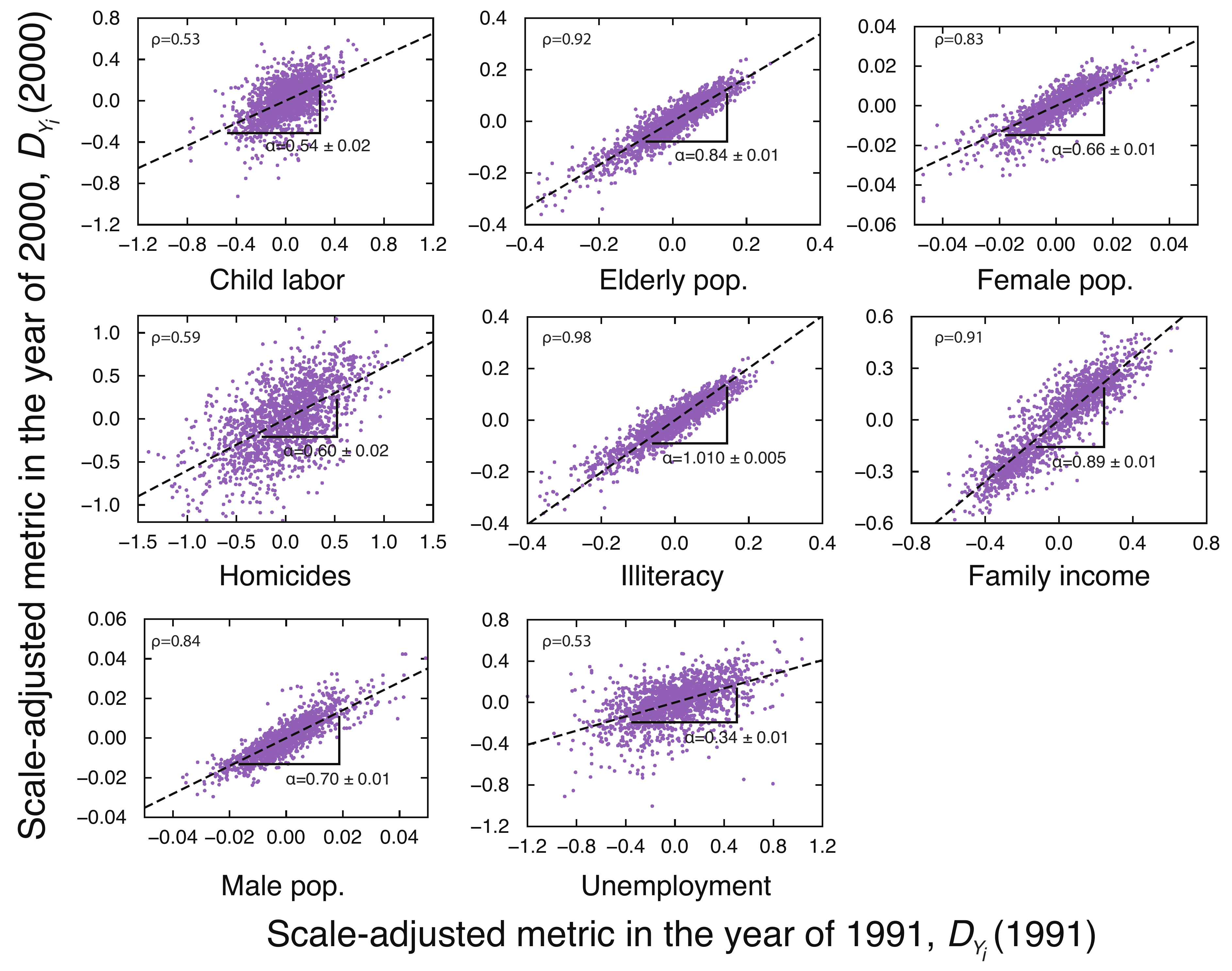}
\end{center}

\noindent {\bf Memory effects in the evolution of the scale-adjusted metrics $D_{Y_i}(t)$.} The purple dots show the values of $D_{Y_i}(2000)$ versus $D_{Y_i}(1991)$ for each city. The dashed lines are fits of the linear model \mbox{$D_{Y_i}(2000) = A_i + \alpha_i \, D_{Y_i}(1991)$} (Eq.~\ref{eq:linmodeldistance}) obtained via ordinary least-square regression. The values of $\alpha_i$ and their standard errors are shown in the plots and also summarized in Table~\ref{tab:2}.
\end{adjustwidth}
\clearpage

\begin{adjustwidth}{-2.25in}{0in}
\subsection*{S4 Fig.}
\label{S4_Fig}
\begin{center}
\includegraphics[scale=0.29]{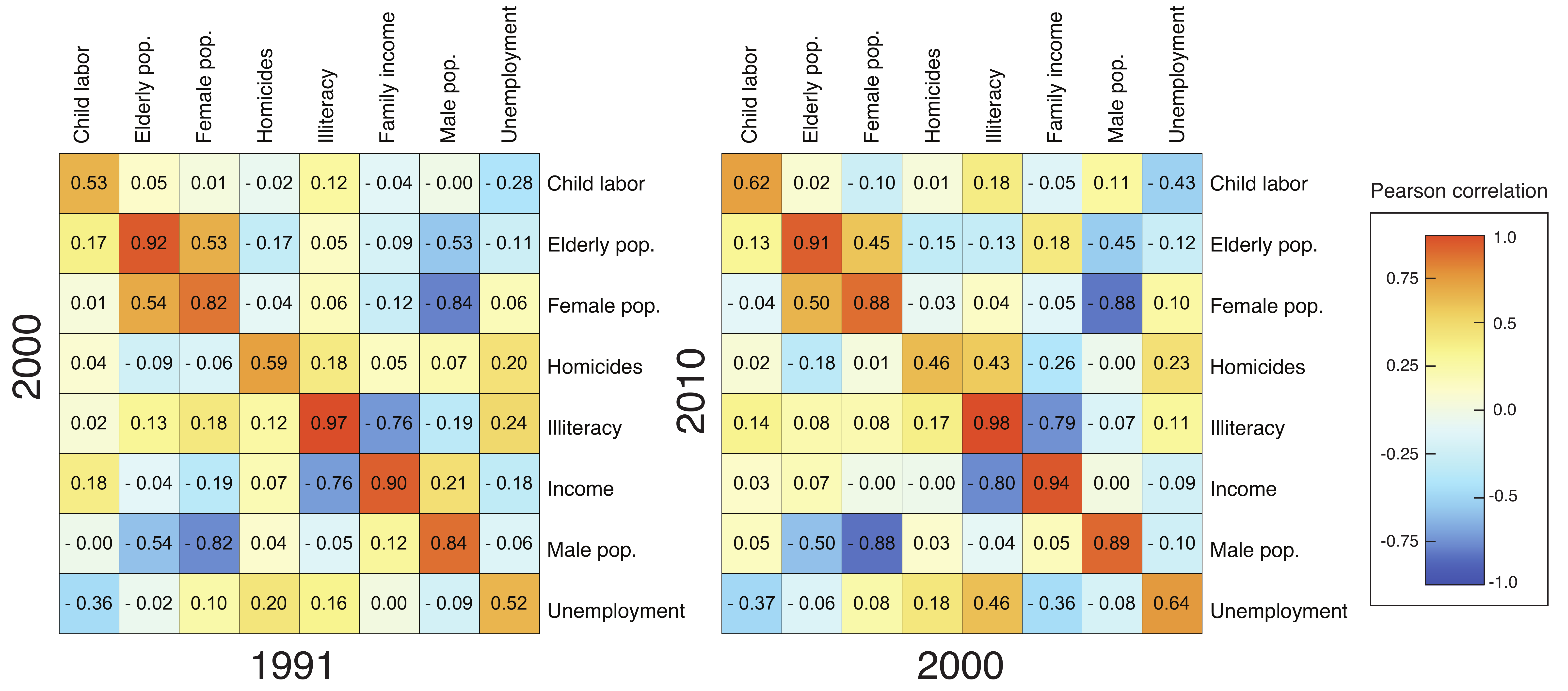}
\end{center}

\noindent {\bf Cross-correlations between the urban indicators.} The matrix plot on left shows the values of the Pearson correlation coefficient between the scale-adjusted metric $D_{Y_i}(t)$ for a given indicator (one indicator per row) in the year $t=2000$ and all the other indicators in the year $t=1991$ (one indicator per column). The right panel does the same for the years $t=2010$ and $t=2000$. The value inside each cell is the Pearson correlation and each one is also colored according to this value. We note that all indicators are strongly correlated with their own past values; furthermore, all indicators also display relevant correlations with at least one other indicator. 
\end{adjustwidth}
\clearpage

\begin{adjustwidth}{-2.25in}{0in}
\subsection*{S5 Fig.}
\label{S5_Fig}
\begin{center}
\includegraphics[scale=0.45]{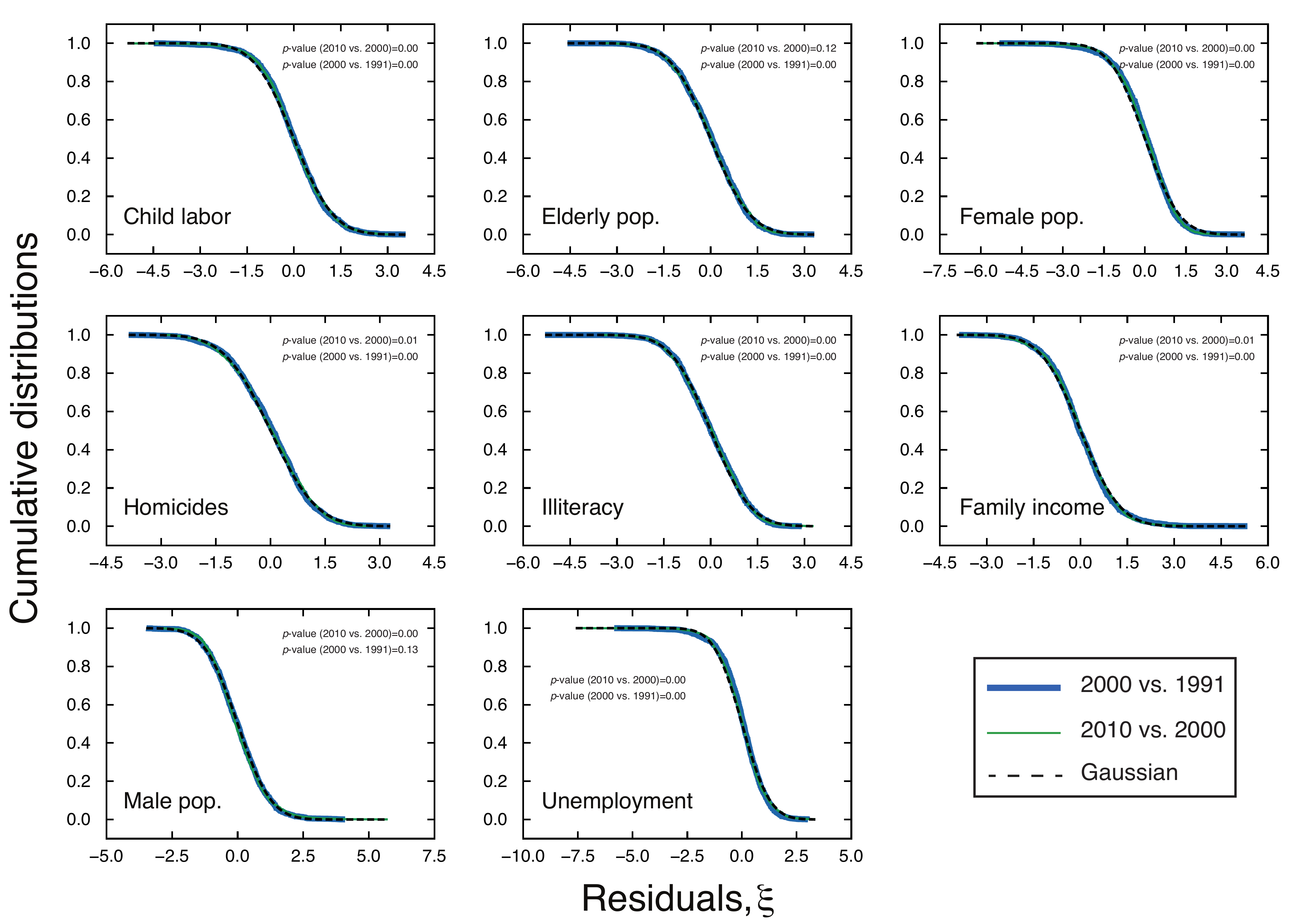}
\end{center}

\noindent {\bf Cumulative distributions of the normalized fluctuations surrounding the relationships between the scale-adjusted metrics $D_{Y_i}(t+\Delta t)$ and $D_{Y_i}(t)$}. The plots show the cumulative distributions of the normalized residuals $\xi$ of the linear regressions between $D_{Y_i}(t+\Delta t)$ and $D_{Y_i}(t)$ (Fig.~\ref{fig:4} and \hyperref[S3_Fig]{S3~Fig.}) for the years 2000-1991 (blue lines) and 2010-2000 (green lines) in comparison with the standard Gaussian (dashed lines). We also show the $p$-values of the Cram\'er von Mises method for testing the null hypotheses that the residuals $\xi$ are normally distributed. We observe that the normality of the data is rejected in most cases (probably due the small heteroskedasticity present in these relationships --- see \hyperref[S6_Fig]{S6~Fig.}). However, no huge differences are observed between the Gaussian cumulative curve and the empirical cumulative distributions, suggesting that $\xi$ can be approximately described as a standard Gaussian noise.
\end{adjustwidth}
\clearpage

\begin{adjustwidth}{-2.25in}{0in}
\subsection*{S6 Fig.}
\label{S6_Fig}
\begin{center}
\includegraphics[scale=0.55]{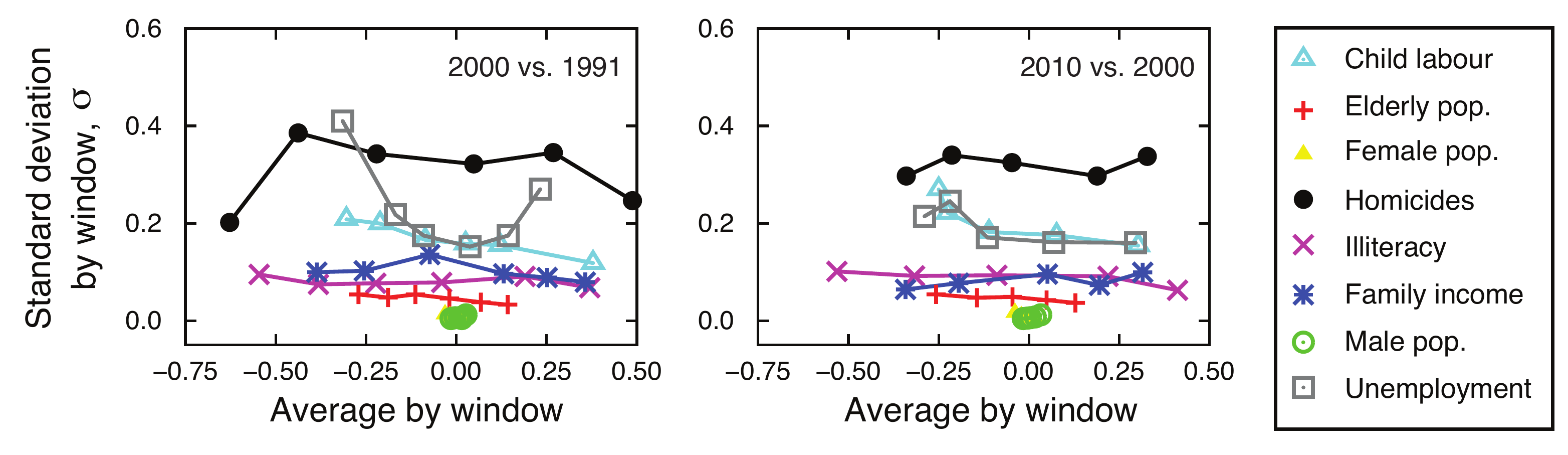}
\end{center}

\noindent {\bf Window-evaluated standard deviation $\sigma$ over the relationship between the scale-adjusted metrics $D_{Y_i}(t+\Delta t)$ and $D_{Y_i}(t)$.} These plots show the standard deviation $\sigma$ of the scale-adjusted metrics $D_{Y_i}(t+\Delta t)$ versus the average value of $D_{Y_i}(t)$ evaluated in five equally spaced windows taken from the relationship between $D_{Y_i}(t+\Delta t)$ and $D_{Y_i}(t)$ (Fig.~\ref{fig:4} and 
\hyperref[S3_Fig]{S3~Fig.}) for the years 2000-1991 (left panel) and 2010-2000 (right panel). We note that the standard deviation can be approximated by a constant for most indicators in both combinations of years. We further observe that the small fluctuations in $\sigma$ are probably the reason of why the Cram\'er von Mises test has rejected the normality of the fluctuations $\xi$ shown in \hyperref[S5_Fig]{S5~Fig.}. When fitting the linear models of Eq.~\ref{eq:glinmodel}, we have also taken into account this small heteroskedasticity (as implemented in the Stata~13 --- \url{http://www.stata.com} --- via the \texttt{robust} option in the \texttt{regress} function) but the linear coefficients remain practically the same.
\end{adjustwidth}
\clearpage

\begin{adjustwidth}{-2.25in}{0in}
\subsection*{S7 Fig.}
\label{S7_Fig}
\begin{center}
\includegraphics[scale=0.37]{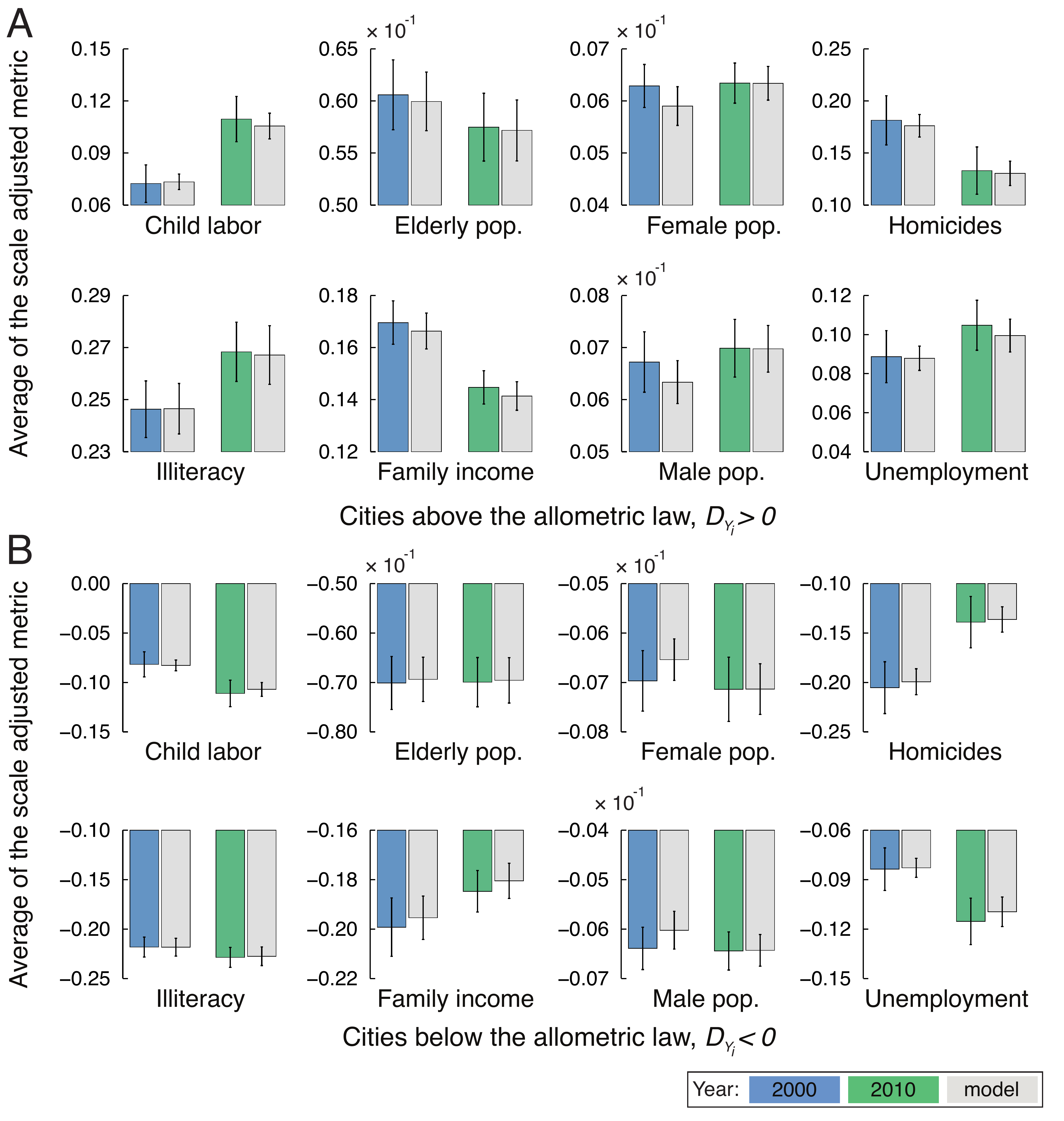}
\end{center}

\noindent {\bf Comparisons between the average values of the scale-adjusted metrics obtained from the linear models and the empirical ones.} We have applied the linear model of Eq.~\ref{eq:glinmodel} for predicting the values of $D_{Y_i}(t)$ in the year of $2000$ only using data from the year of $1991$ as well as for predicting $D_{Y_i}(t)$ in the year of $2010$ only using data from the year of $2000$. In both cases, we have calculated the average $D_{Y_i}(t)$ for the predictions (gray bars) after grouping the cities in above (A) and below (B) the allometric laws (in that year) and compared these results with the same averages evaluated using the empirical data (blue bars for year of 2000 and green bars for the year of 2010). The errors bars are 95\% bootstrapping confidence intervals for the average values. We observe that the predicted average values are in very good agreement with the empirical values for all urban indicators in both years.
\end{adjustwidth}
\clearpage

\begin{adjustwidth}{-2.25in}{0in}
\subsection*{S8 Fig.}
\label{S8_Fig}
\begin{center}
\includegraphics[scale=0.5]{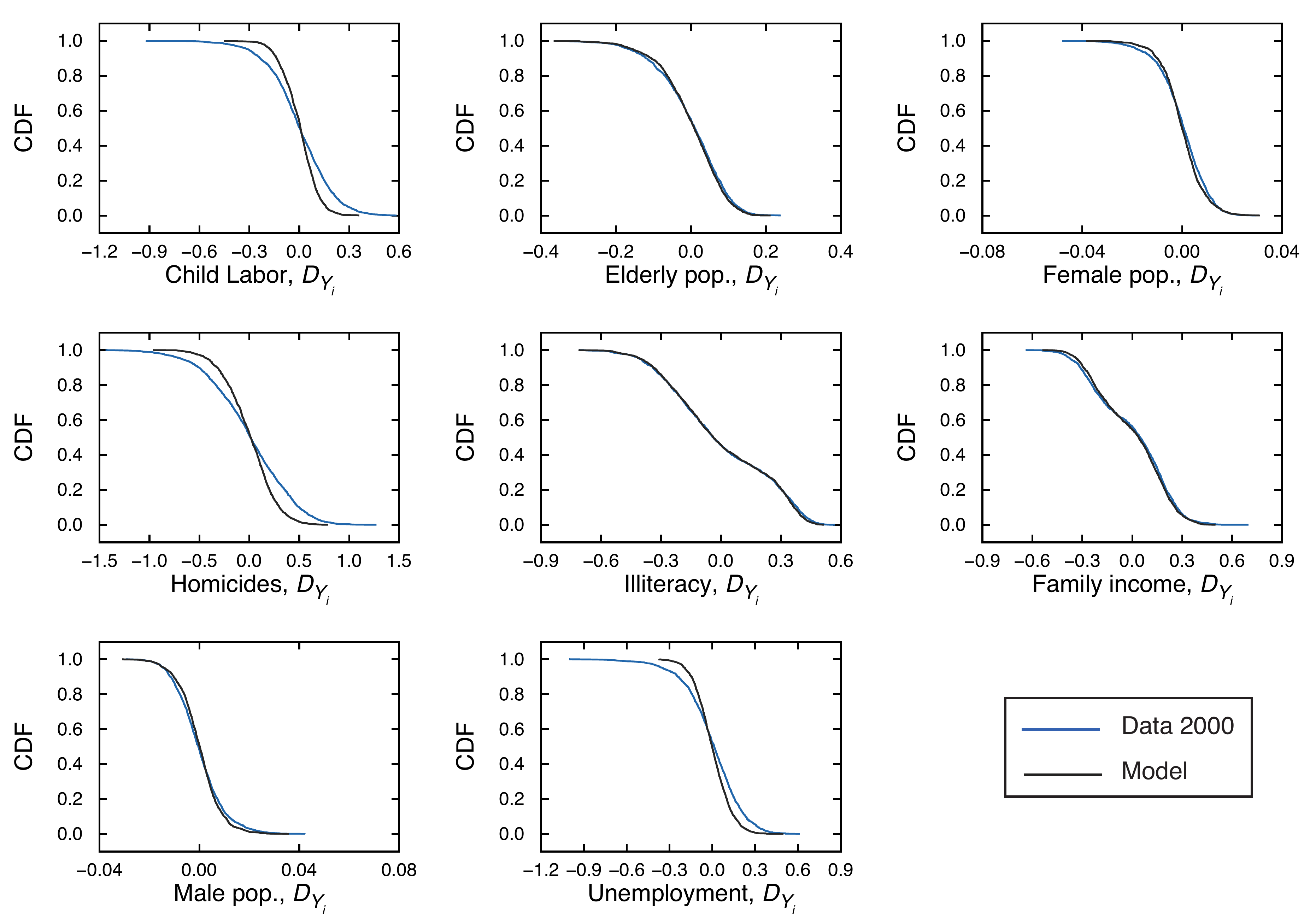}
\end{center}

\noindent {\bf Comparisons between the cumulative distributions of the scale-adjusted metrics obtained from the linear models and the empirical ones.} We have obtained the values of $D_{Y_i}(t)$ in the year of $2000$ using the linear model of Eq.~\ref{eq:glinmodel} and by employing data from the year of $1991$. We thus calculated the cumulative distributions functions (CDF) of $D_{Y_i}(t)$ for the predicted values (black lines) and the empirical ones (blue lines). We observe that the agreement is very good for the population indicators, illiteracy and family income; for the other indicators we observe that the model fails in reproducing the tails of the distributions.
\end{adjustwidth}
\clearpage

\begin{adjustwidth}{-2.25in}{0in}
\subsection*{S9 Fig.}
\label{S9_Fig}
\begin{center}
\includegraphics[scale=0.5]{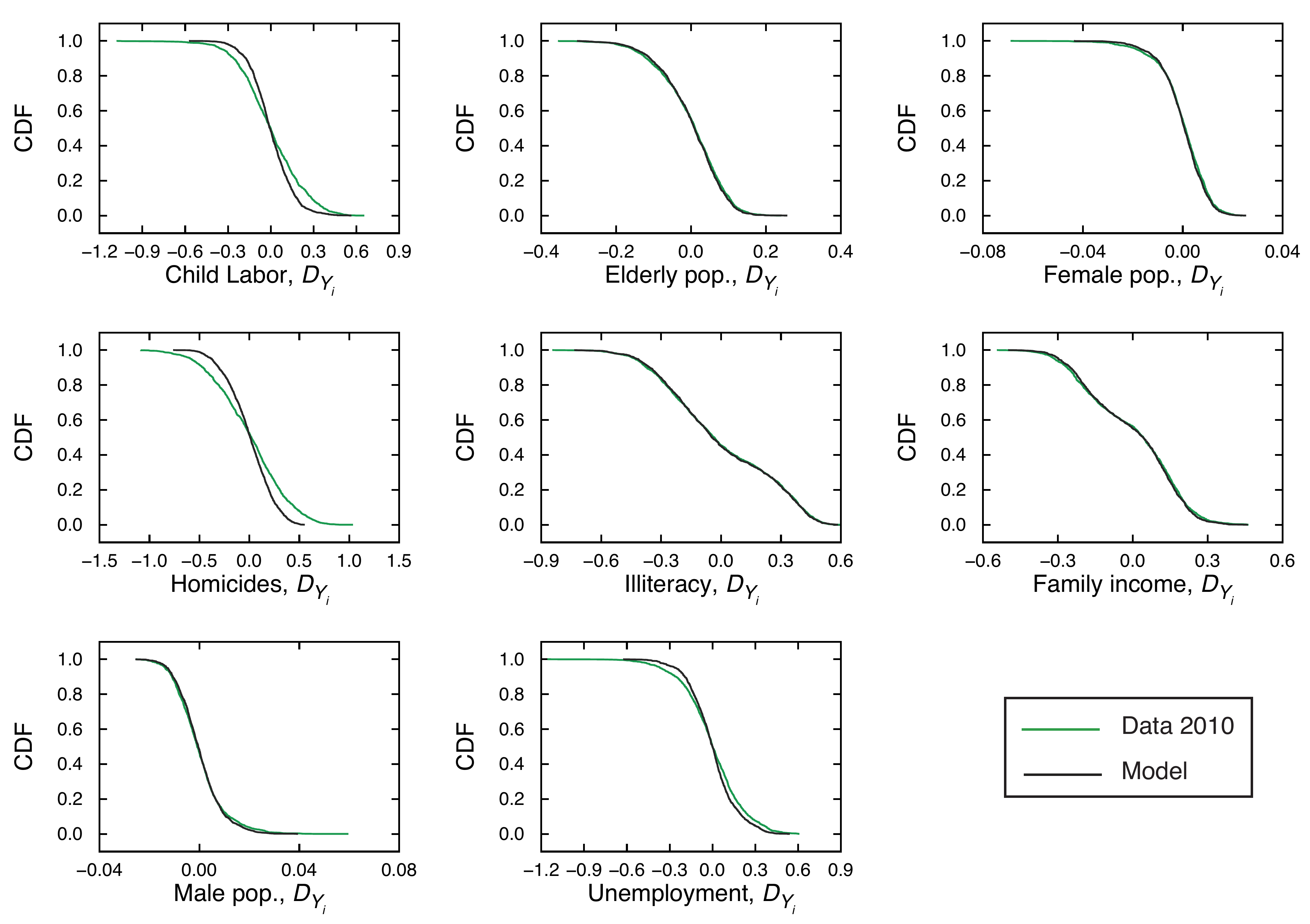}
\end{center}

\noindent {\bf Comparisons between the cumulative distributions of the scale-adjusted metrics obtained from the linear models and the empirical ones.} The same as \hyperref[S8_Fig]{S8~Fig.} considering data from the year of $2010$.
\end{adjustwidth}
\clearpage

\subsection*{S1 Text}
\label{S1_Text}
{\bf Coefficients of the linear regression model of the Eq.~\ref{eq:glinmodel}}. Values of the linear coefficients in the model of the Eq.~\ref{eq:glinmodel} for the relationships $D_{Y_i}(2000)$ versus $D_{Y_i}(1991)$ and $D_{Y_i}(2010)$ versus $D_{Y_i}(2000)$ for the eight urban indicators.

\end{document}